# THE OBSCURED ACTIVE NUCLEUS OF NGC 7172 AS SEEN BY NuSTAR

## Vasylenko A. A.

*Main astronomical observatory of NAS of Ukraine, Kyiv, 03143, UKRAINE*

*e-mail: vasylenko_a@mao.kiev.ua*

We present the hard X-ray spectral analysis of NGC 7172, the nearby (z=0.0087) Seyfert 2 galaxy. This analysis is based on the spectral data from a 32 ks *NuSTAR* observation conducted in 2014 (ID 6006130800). The *NuSTAR* 3-64 keV spectrum of the source showed a constant Compton-thin obscuration $N_H \approx 8 \cdot 10^{22}$ cm$^{-2}$, which is similar to that observed by *XMM-Newton*, *Suzaku*, *ASCA* and *BeppoSAX* over past 30 years.

We revealed the presence of a primary power-law continuum with $\Gamma \approx 1.8$, a moderate reflection component with <R> ~ 0.44 (adopting the *ad-hoc* disk-like reflection model `pexmon` (Nandra et al., 2007)) and a narrow Fe K$_\alpha$ line with $EW = 67^{+13}_{-14}$ eV. The application of the numerical torus models, such as `BNTorus` (Brightman & Nandra, 2011) and `MYTorus` (Murphy & Yaqoob, 2009), confirmed the Compton-thin type of the Seyfert nucleus and allowed us to obtain an estimations of the torus opening angle $\Theta_t$ ~ 59° and inclination $\Theta_i$ ~ 61°. Interestingly, that additional reflection component with R≈0.35 is needed unexpectedly during the fit with `BNTorus` model, thus this model is likely to be inappropriate for NGC 7172.

The measured (using `MYTorus` model) intrinsic 2-10 keV ($L_{intr}$(2-10 keV)=(1.14-1.23)·10$^{43}$ erg/s) and 10-40 keV ($L_{intr}$(10-40 keV)=(1.56-1.62)·10$^{43}$ erg/s) luminosity of NGC 7172 indicate the brightening of source in X-rays as compare with previous data for ~18 years. Using the data of previous observations, we demonstrate also the long-term variability of $L_{intr}$(2-10 keV) almost by order and $EW$ FeK$_\alpha$ by factor ~4 on a timescale of ~12 years. It coincides to the distance of d~3.7 pc between central source and reprocessing medium. Such changes of intrinsic luminosity without changes in $N_H$ value indicate the variability of a central source. In the same time the results of spectral analysis with the presence of the lag between variability of intrinsic luminosity $L_{intr}$(2-10 keV) and $EW$ FeK$_\alpha$, as well as a behavior of intensity $I_{FeK\alpha}$ of the line, are in agreement with the scenario, where the observed FeK$_\alpha$ line is generated in a distant gas-dust torus.

Keywords: galaxies: active --- X-rays: galaxies --- X-rays: individual (NGC 7172)

# Затенённое активное ядро галактики NGC 7172 по наблюдению NuSTAR

А.А. Василенко

Главная астрономическая обсерватория НАН Украины, ул. Академика Заболотного 27, Киев, 03143

*Проанализированы свойства рентгеновского излучения активного ядра галактики NGC 7172 типа Сейферт 2 по данным наблюдения космической обсерватории NuSTAR в 2014 году. Источник демонстрирует постоянное во времени поглощение $N_H \approx 8 \cdot 10^{22}$ см$^{-2}$, значение которого сравнимо с полученным поглощением из наблюдений XMM-Newton, Suzaku, ASCA и BeppoSAX за предыдущие ~30 лет. Базовый спектральный анализ выявил присутствие умеренной компоненты отражения с <R> ~ 0.44 и узкой линии Fe K$_\alpha$ с $EW = 67^{+13}_{-14}$ эВ. Мы применили численные модели тора, которые подтвердили Компотоновски-тонкий тип сейфертовского ядра галактики, а также помогли получить оценку угла наклона тора $\Theta_i$ ~ 61˚ и его раскрытия $\Theta_t$ ~ 59˚. Мы также продемонстрировали, что данные предыдущих наблюдений показывают изменчивость собственной светимости $L_{intr}$(2-10 кэВ) примерно на порядок и эквивалентной ширины EW FeK$_\alpha$ в 4 раза на интервале ~12 лет, что соответствует расстоянию d~3.7 пк. Такое поведение $L_{intr}$(2-10 кэВ) свидетельствует о переменности центрального источника. Наряду с этим, результаты спектрального анализа вместе с задержкой между изменением светимости и EW FeK$_\alpha$, а также изменение интенсивности $I_{FeK\alpha}$, лучше всего соответствуют варианту рождения линии FeK$_\alpha$ в отдалённом газопылевом торе.*

ЗАТЕМНЕНЕ АКТИВНЕ ЯДРО ГАЛАКТИКИ NGC 7172 ЗА СПОСТЕРЕЖЕННЯМ NuSTAR, *Василенко А. А.* — *Проаналізовано властивості рентгенівського випромінювання активного ядра галактики NGC 7172 типу Сейферт 2 за даними спостереження космічної обсерваторії NuSTAR в 2014 році. Джерело демонструє постійне в часі поглинання $N_H \approx 8 \cdot 10^{22}$ см$^{-2}$, величина якого порівняна з отриманим поглинанням зі спостережень XMM-Newton, Suzaku, ASCA та BeppoSAX за попередні ~30 років. Базовий спектральний аналіз виявив присутність помірної компоненти відбиття з <R>~0.44 та вузької лінії Fe K$_\alpha$ з $EW = 67^{+13}_{-14}$ eВ. Ми застосували чисельні моделі тору, які підтвердили Компотонівськи-тонкий тип сейфертівського ядра галактики, а також допомогли отримати оцінку куту нахилу тору $\Theta_i$ ~ 61˚ та його розкриття $\Theta_t$ ~ 59˚. Ми також продемонстрували, що дані попередніх спостережень показують змінність власної світності $L_{intr}$(2-10 кеВ) приблизно на порядок та еквівалентної ширини EW FeK$_\alpha$ в 4 рази на інтервалі ~12 років, що відповідає відстані d~3.7 пк. Така*

*поведінка $L_{intr}$(2-10 кеВ) свідчить про змінну поведінку центрального джерела. Водночас результати спектрального аналізу разом із затримкою між зміною світності та EW $FeK_\alpha$, а також зміна інтенсивності $I_{FeK\alpha}$, найкраще відповідають варіанту народження лінії $FeK_\alpha$ у віддаленому газопиловому торі.*

## 1 Введение

Галактика NGC 7172 является видимой почти с ребра галактикой раннего типа, которая входит в компактную группу галактик HCG 90. Оптически классифицированная как Сейферт 2 [30], эта галактика находится на расстоянии z=0.0087.

NGC 7172 наблюдалась в рентгеновском диапазоне почти всеми основными миссиями. Первое наблюдение было предпринято спутником *EXOSAT* в диапазоне 2-10 кэВ, которое показало наличие степенного спектра со степенным индексом Γ=1.84 и поглощением (т.е. столбцевой концентрацией водорода) $N_H \approx 10^{23}$ см$^{-2}$ [33]. Анализ двух наблюдений с помощью *ASCA* [12,29,34] показал наличие меньшего наклона континуума Γ≈1.5 при значении поглощения $N_H \approx 8 \cdot 10^{22}$ см$^{-2}$, а также присутствие эмиссионной линии железа Fe $K_\alpha$ 6.4 кэВ. Кроме того, в [12] была обнаружена переменность кривой блеска на уровне 30% в пределах нескольких часов. В работах [1,7,28] представлен анализ двух широкодиапазонных (1.65-50 кэВ) наблюдений с помощью *BeppoSAX*, в которых определены значения степенного индекса в пределах 1.6-1.9 в зависимости от модели спектра, значения поглощения в пределах 8.3-11·$10^{22}$ см$^{-2}$, а также присутствие узкой линии Fe $K_\alpha$ 6.4 кэВ. Было отмечено наличие компоненты отражения от нейтральной среды [28]. Обсерватория *XMM-Newton* трижды наблюдала NGC 7172 и все наблюдения показывают практически одинаковые значения степенного индекса Γ≈1.6 и поглощения ~7-8·$10^{22}$ см$^{-2}$ [напр., 3,8,13,17,36] (в работе [11], где применялась модель с двумя компонентами отражения, $N_H$ ~1.3·$10^{23}$ см$^{-2}$), а также наличие узкой линии Fe $K_\alpha$. В работе [11] проверка на наличие релятивистского размытия для линии железа показала отсутствие такового. Анализ наблюдения Suzaku в широком диапазоне 0,5-150 кэВ [9,16] показал значение степенного индекса Γ≈1.7, поглощения ~8.9·$10^{22}$ см$^{-2}$, наличие узкой линии Fe $K_\alpha$, а также отсутствие в спектре релятивистских эффектов.

Рентгеновскому излучению от активного ядра NGC 7172 присуща кратковременная переменность, которая, например, была детально изучена в работе [3], а также долговременная переменность (см., например [12,13]). При этом отсутствует значительное изменение величины поглощения (в зависимости от модели 7-9·$10^{22}$ см$^{-2}$), по значению которой галактика относится к Комптоновски-тонкой по поглощению.

Наличие и вклад нейтрального отражения в рентгеновский спектр NGC 7172 не до конца изучен, несмотря на присутствие линии Fe $K_\alpha$, и не рассматривается как необходимая в ряде работ [1,12,13,17]. В то же время, в работах [3,7,11,16,28,34] авторы приходят к выводу о его присутствии в спектре.

В данной работе представлен спектральный анализ рентгеновского спектра NGC 7172, полученного космическим аппаратом миссии *NuSTAR* (Nuclear Spectroscopic Telescope Array), который наблюдает в диапазоне энергий 3-79 кэВ. Благодаря высокой чувствительности обсерватории, качество её данных позволило получить характеристики спектра отражения, а также протестировать несколько моделей структуры поглощающей среды.

Статья структурирована следующим образом: в п. 2 описана процедура обработки сырых данных, в п. 3 — краткое описание кривой блеска, в п. 4 представлен спектральный анализ, а в п. 5 обсуждается интерпретация результатов и формулируются соответствующие выводы.

**2 Обработка данных**

Галактика NGC 7172 наблюдалась с помощью NuSTAR 07.10.2014 (ID 60061308002) продолжительностью 32 тыс. сек. Исходные данные были обработаны с помощью программ пакета NuSTARDAS v.1.6.0 (NuSTAR Data Analysis Software package). Калиброванные и очищенные файлы событий были получены с использованием калибровочных файлов *NuSTAR* CalDB (20171204) и стандартных критериев отбора в подпрограмме *nupipeline*. Для получения спектров источника и фона, а также очищенных кривых блеска была использована подпрограмма *nuproducts*. Области источника и фона были выбраны в обоих детекторах FPMA и FPMB как круговые области радиусом 60″ и 70″ соответственно (для фона — в области без других источников). Для минимизации систематических эффектов, спектры, полученные камерами FPMA и FPMB, не были объединены в один, хотя их подгонка была одновременной.

**3 Кривая блеска**

На Рис. 1 приведена исправленная на фон кривая блеска *NuSTAR* FPMA+FPMB в диапазонах 3-10 кэВ и 10-60 кэВ, а также их отношение. Данные сгруппированы в бины с шириной интервала 900 с. Для анализа кривой блеска была использована программа FTOOLS *lcstats*. Средние значения скорости счета со стандартными отклонениями имеют значения $3.12\pm0.27$ с$^{-1}$ для 3-10 кэВ и $1.56\pm0.19$ с$^{-1}$ для 10-60 кэВ соответственно. Принимая гипотезу об отсутствии вариаций кривых блеска в обоих диапазонах, соответственно получим

$\chi^2$/d.o.f.=117.3/49 и $\chi^2$/d.o.f.=96.27/49, что свидетельствует о присутствии умеренной кратковременной переменности. Вследствие отсутствия значительных по амплитуде вариаций в кривых блеска, в дальнейшем используется усреднённый по времени спектр.

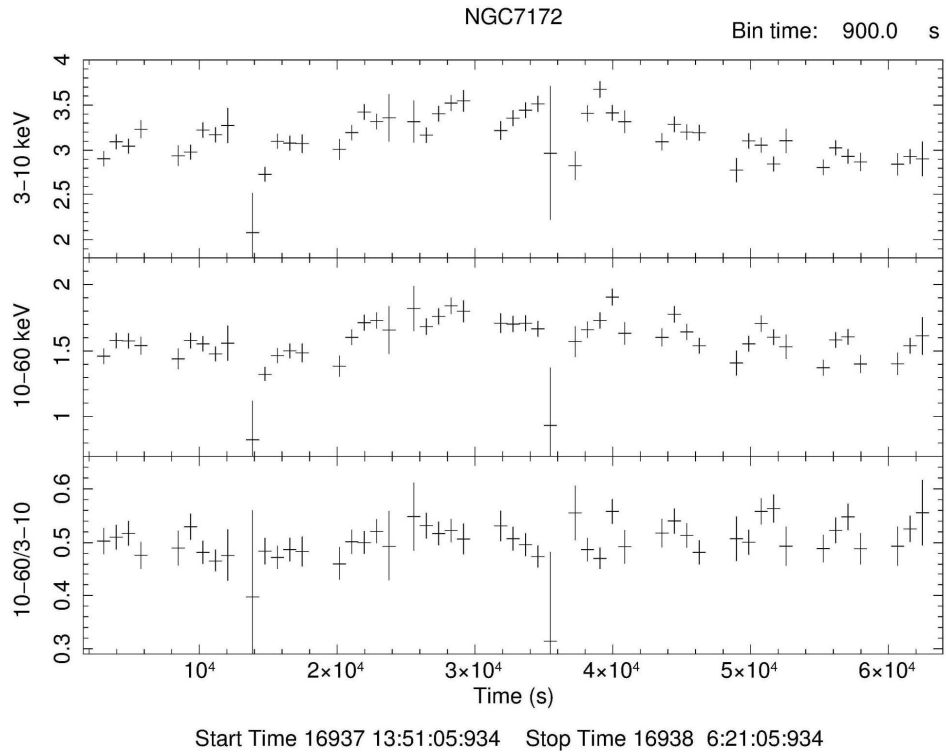

*Рис. 1*. Кривые блеска FPMA+FPMB в диапазонах 3-10 кэВ (вверху), 10-60 кэВ (посередине), а также их отношение (внизу).

## 4 Спектральный анализ

Анализ спектра производился при помощи специализированной программы XSPEC v.12.9.0u, которая является частью программного пакета HEASOFT v.6.19. Ошибки параметров, приведённые в данной работе, отображают 90% доверительный интервал для одного параметра ($\Delta\chi$=2.71). При вычислении светимостей были использованы космологические параметры $H_0$=70 км с$^{-1}$ Мпс$^{-1}$, $\Lambda_0$=0.73, $\Omega_M$=0.27 [5]. Детекторы FPMA/FPMB могут получать спектры вплоть до 79 кэВ, но в нашем случае диапазон энергий был ограничен до ~ 60 кэВ из-за значительного фона на высших энергиях. Таким образом, в спектральном анализе используется диапазон энергий 3-64 кэВ.

Величина поглощения в Галактике определяется столбцовой концентрацией $N_{H,gal}$ = 1.9·10$^{20}$ см$^{-2}$ согласно [15] и учитывалась моделью tbabs [37]. Для учёта расхождений во взаимной калибровке камер FPMA и FPMB была введена постоянная интеркалибровки *C* (в моделях обозначена как `constant`), которая в процессе подгонки равнялась 0.98+/-0.01, т.е. не более 5%, что соответствует [20].

**Феноменологические модели.** Для начала мы получили параметры континуума, для чего был исключён интервал энергий в диапазоне 5.5 – 7.5 кэВ, где возможен существенный вклад эмиссионных линий, и проведена подгонка спектра со степенным энергетическим распределением, где нормировка, фотонный индекс Γ и значение внутреннего поглощения $N_H$ являются свободными входными параметрами модели. Начальная базовая модель имела вид `Tbabs*zTbabs*zpo*constant`. Была получена неплохая подгонка ($\chi^2$/d.o.f.=1093/1077) со значениями параметров Γ=1.67±0.02 и $N_H$=(7.7±0.3)·$10^{22}$ см$^{-2}$. Экстраполяция на диапазон 5.5 – 7.5 кэВ показывает хорошо заметную эмиссионную линию, что отражается на значении статистики $\chi^2$/d.o.f.=1299/1079. Включение в модель линии с гауссовским профилем `zgauss` значительно улучшило статистику ($\chi^2$/d.o.f.=1140/1125) и показало параметры линии $E_{line}$=6.29±0.07 кэВ, $\sigma = 190^{+165}_{-115}$ эВ. Несмотря на хорошее значение статистики, в спектре наблюдается небольшой горб в диапазоне 20-40 кэВ, соответствующий по энергии т.н. «комптоновскому горбу», а также порог поглощения около 7 кэВ, которые являются характерными признаками наличия компоненты отражения. Для её учёта, а также для учёта возможного экспоненциального высокоэнергетического обрезания, мы добавили компоненту нейтрального комптоновского отражения `pexrav` [21] и заменили простой степенной закон на такой же, но с энергией обрезания $E_{cut}$. Модель `pexrav` включает в себя параметр относительного отражения R, который определяется как отношение телесного угла аккреционного диска (в виде плоской непрозрачной пластины), под которым диск наблюдается из первичного источника (т.н. короны диска), к полусфере 2π. Таким образом, окончательная базовая модель выглядит как: `Tbabs*(zTbabs*cutoffpl+pexrav+zgauss)*constant`. Получены значения Γ=1.83±0.05, $N_H$=(8.9±0.5)·$10^{22}$ см$^{-2}$, R=0.50±0.13 при $\chi^2$/d.o.f.= 1106/1124. Параметры линии $E_{line}$=6.33±0.06 кэВ, $\sigma = 93^{+100}_{-90}$ эВ, эквивалентная ширина $EW = 67^{+13}_{-14}$ эВ. Значение энергии обрезания не было получено и поэтому было зафиксировано на $E_{cut}$=500 кэВ. Результирующая подгонка рентгеновского спектра галактики NGC 7172 показана на Рис. 3, значения параметров приведены в Табл. 1.

Полученная энергия линии $E_{line}$~6.33 кэВ (Рис. 2) может быть интерпретирована как смещённая линия Fe $K_\alpha$. В то же время в работах [3,9,17] были получены значения, близкие к традиционной величине $E_{line}$≈6.4 кэВ. Если мы зафиксируем энергию линии на 6.4 кэВ, то значения ширины σ и ширины EW не изменяются в границах ошибок. Таким образом, смещённая величина $E_{line}$ может быть объяснена худшим энергетическим разрешением матриц детекторов FPMA/FPMB (FWHM~400 эВ) на 6 кэВ по сравнению с детекторами камер XMM-

Newton/EPIC и Suzaku/XIS (FWHM~150 эВ), данные которых использовались в упомянутых работах.

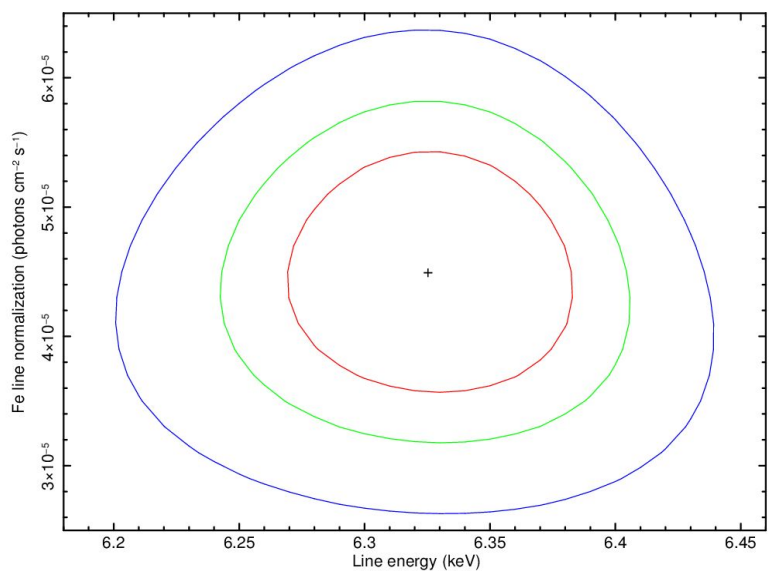

*Рис. 2*. Контуры доверительных интервалов для энергии линии Fe K$_\alpha$ и ее нормировки. Показаны контуры 68.3%, 90% и 99%.

Поскольку нейтральная линия железа Fe K$_\alpha$ и «комптоновский горб» являются проявлениями одной спектральной компоненты отражения, единая модель, которая описывает эти части вместе, может дать лучшие значения параметров. Поэтому мы заменили модели `pexrav` и `zgauss` на одну модель `pexmon` [25], которая самосогласованно включает линию Fe K$_\alpha$ и «комптоновский горб» (`Tbabs*(zTbabs*cutoffpl+gsmooth*pexmon)*constant`). Компонента `gsmooth` учитывает уширения эмиссионных линий с гауссовским профилем (свободный параметр *σ*) Значения основных параметров Γ=1.80±0.02, $N_H$=(8.6±0.3)·$10^{22}$ см$^{-2}$, R=0.44±0.04 при $\chi^2$/d.o.f.=1104/1124.

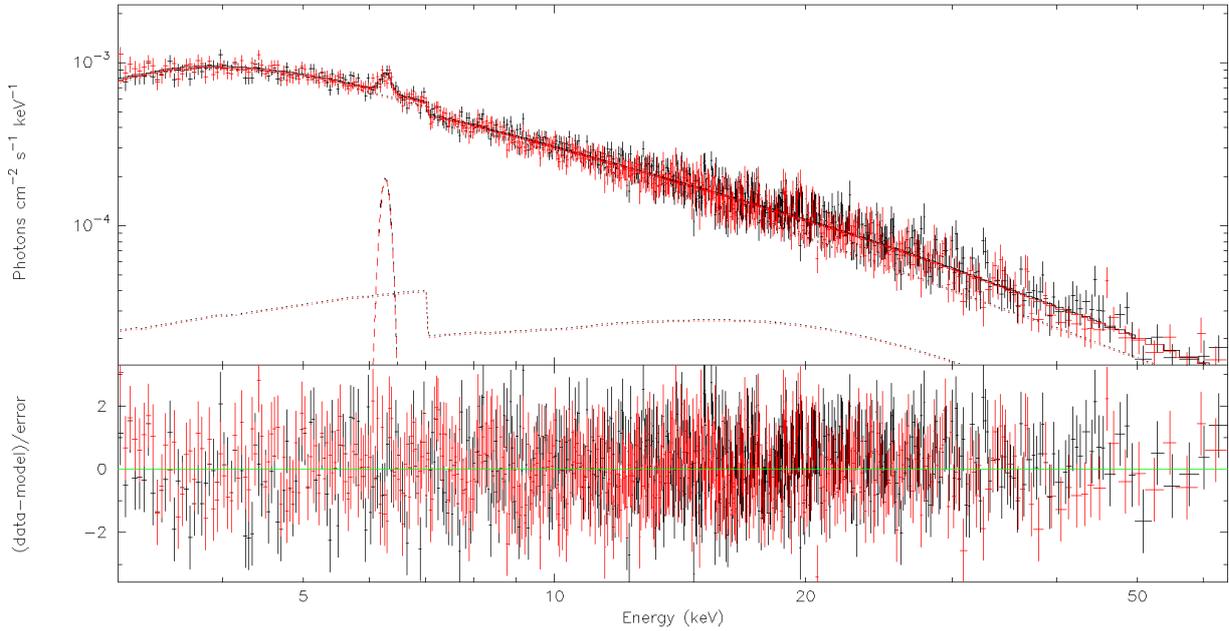

*Рис. 3.* Лучшая подгонка спектра с использованием базовой модели `pexrav`. Нижняя панель – остаточные отклонения. Сплошная кривая – суммарная модель, точечный пунктир – отдельные вклады степенного континуума и отражения, пунктир – линия Fe $K_\alpha$.

Для лучшего учёта поглощения излучения от центрального источника, вместо простого степенного закона с дополнительным поглощением, была также использована более физическая модель `plcabs` [39]. Эта модель описывает континуум спектра рентгеновского излучения от изотропного источника в центре сферической формы поглощающего материала с учётом комптоновского рассеяния и поглощения. Выражение для полной модели `Tbabs*(plcabs+gsmooth*pexmon)*constant`. В этом случае также была получена очень хорошая подгонка $\chi^2$/d.o.f.=1109/1124 и следующие значения спектральных параметров $\Gamma$=1.79±0.02, $N_H$=(8.2±0.4)·$10^{22}$ см$^{-2}$, R=0.41±0.05.

**Численные модели газопылевого тора.** Для изучения структуры поглощающей среды и более реалистичного учёта её взаимодействия с рентгеновским излучением от центрального источника, была предпринята подгонка спектра с использованием физических численных моделей, полученных из Монте-Карло моделирования, а именно с табличными моделями `BNTorus` [6] и `MYTorus` [23,40,41].

Модель `BNTorus` описывает поглощающий материал в форме сферического тора с изменяемым углом раскрытия $\Theta_t$ полярных конусов от 25.8° до 84.3°, а также углом наклона экватора тора $\Theta_i$ от 18.2° до 87.1°. Поглощение на луче зрения совпадает с поглощением в экваториальной плоскости и не зависит от угла наклона. Модель самосогласованно содержит в себе компоненты пройденного, рассеянного и отражённого излучения, а также включает в себя эмиссионные линии Fe K$_\alpha$, Fe K$_\beta$, Ni K$_\alpha$ и ряда других элементов в мягком рентгене. По отдельности компоненты не разделяются.

В начале углы наклона были зафиксированы, — на верхнем значении $\Theta_i$=87.1° для наклона, и на нижнем пределе $\Theta_t$=25.8° для раскрытия, так как они не определяются при одновременной вариации со степенным индексом. Полученная величина поглощения равна $N_H$=(7.5±0.2)·10$^{22}$ см$^{-2}$, степенной индекс Γ=1.73±0.01. Несмотря на то, что подгонка показывает хорошую статистику $\chi^2$/d.o.f.=1186/1126, на спектре чётко выделяется недооценка уровня потока между 6-7 кэВ, то есть в области линии железа Fe K$_\alpha$, а также в диапазоне 20-40 кэВ, который соответствует области «комптоновского горба». Эти особенности свидетельствуют о необходимости дополнительной модели отражения, для чего в модель спектра была добавлена компонента pexmon. Выражение для полной модели стало иметь вид:

**Модель T** = Tbabs*gsmooth*(atable{torus1006.fits}+pexmon)*constant.

Результирующая подгонка показывает очень хорошую статистику $\chi^2$/d.o.f.=1116/1125 изменившись на $\Delta\chi^2$=70 для 1 d.o.f. Применение теста Фишера показывает величину $F_{value}$=70.22 и соответствующую вероятность $p$=1.6·10$^{-16}$, что свидетельствует о статистической обоснованности добавления модели отражения. Полученная величина поглощения $N_H$=(7.7±0.2)·10$^{22}$ см$^{-2}$, степенной индекс стал более «мягким» Γ=1.80±0.01, а параметр отражения R=0.35±0.04. Величина R немного меньше полученных значений при базовом моделировании, что ожидаемо, поскольку модель BNTorus уже включает в себя компоненту рассеянного и отражённого излучения, но её присутствие в принципе является несколько неожиданным[1]. Также удалось получить оценку угла раскрытия газопылевого тора $\Theta_t=(59^{+16}_{-20})°$. Величина угла наклона определяется более грубо $\Theta_i=75^{+u}_{-12}$ град[2].

Вторая численная модель — модель MYTorus, описывает поглощающий материал с тороподобной геометрией с фиксированным углом раскрытия 60° (фактор перекрытия ≡ 0.5), изменяемым углом наклона и включает в себя несколько компонентов. Первая и основная из них (MYTZ) отвечает за модификацию первичного излучения, прошедшего сквозь газопылевой тор. Вторая компонента (MYTS) представляет собой отражённое и рассеянное в торе излучение первичного источника. Эти две компоненты континуума дополняет третья компонента (MYTL), которая описывает излучение в эмиссионных линиях Fe K$_\alpha$, Fe K$_\beta$ и Ni K$_\alpha$, которые генерируются в этом же торе, т.е. эта компонента является согласованной с параметрами континуума. Для описания первичного источника рентгеновского излучения был выбран степенной закон. Для учёта экспоненциального энергетического обрезания, была выбрана табличная модель[3] с наибольшим значением 500 кэВ, поскольку точное значение $E_{cut}$ не было определено в базовой подгонке.

---

[1] Дополнительная компонента отражения, возможно, обуславливается тем, что а) геометрия модели BNTorus не подходит к описанию данного спектра, или б) отражённое от дальней стенки тора излучение в модели BNTorus при любых углах наклона и раскрытия считается таким, что не претерпевает поглощения или рассеяния при дальнейшем прохождении ближней стенки тора.
[2] "u" - величина не определена

Мы использовали стандартный вариант модели MYTorus – т.н. "coupled" режим[4], при котором все параметры компонент MYTS и MYTL были приравнены к параметрам компоненты первичного излучения MYTZ. Константы относительной нормировки $A_i$ между всеми тремя компонентами (т.е. $A_S$, $A_L$ и $A_Z$) были зафиксированы и равнялись 1, что точно соответствует оригинальному варианту модели MYTorus. Значения поглощения между всеми тремя компонентами также приравнивались ($N_{H(MYTZ)} = N_{H(MYTS)} = N_{H(MYTL)} = N_{H(eq)}$), что соответствует стандартному однородному тору. Свободными параметрами являются индекс Γ и нормировка степенного закона, поглощение в экваториальной плоскости $N_{H(eq)}$ и угол наклона газопылевого тора $\Theta_i$. Таким образом, выражение для полной модели имеет вид:

**Модель M =**
```
constant*Tbabs*(zpowerlw*etable{mytorus_Ezero_v00.fits}
+constant*atable{mytorus_scatteredH500_v00.fits}
+constant*(gsmooth*atable{mytl_V000010nEp000H500_v00.fits})).
```

Применение модели показало хорошую подгонку $\chi^2$/d.o.f.=1136/1125, степенной индекс Γ=1.71±0.01 и угол наклона $\Theta_i$≈60.4, что близко к касательному углу. Неожиданно было получено большое значение поглощения $N_{H(eq)}$=(7.5±1.7)·$10^{23}$ см$^{-2}$. Поглощения на луче зрения в "coupled" режиме модели MYTorus может быть получено с помощью уравнения (см. раздел 3.1 в [23]):

$$N_{H(l.o.s)} = N_{H(eq)}(1 - 4\cos^2\Theta_i)^{1/2},$$

откуда следует, что $N_{H(l.o.s)}$≈1,2·$10^{23}$ см$^{-2}$, что приблизительно на 60% больше $N_{H(l.o.s)}$, полученных в данной и других работах. Поэтому, как следующий шаг, мы «отвязали» значение нормировок $A_S$ и $A_L$ от первичной компоненты, позволив им изменяться, но с условием, что $A_S=A_L$, что подразумевает общий регион формирования рассеянного излучения и эмиссионных линий. Важно отметить, что, как подчёркивается в [40], величина $A_S$ не может быть напрямую интерпретирована как некий аналог фактора перекрытия, поскольку точная форма рассеянного континуума варьируется с изменением этого же фактора. В результате была также получена хорошая подгонка с $\chi^2$/d.o.f.=1143/1125, значением степенного индекса Γ=1.71±0.01 и аналогичным углом наклона $\Theta_i = (61.1^{+0.9}_{-0.4})°$. При этом, небольшое значение константы $A_S = A_L = 1.20 \pm 0.07$ привело к более правдоподобному поглощению $N_{H(l.o.s)} \approx 8 \cdot 10^{22}$ см$^{-2}$ из полученного на экваторе тора $N_{H(eq)} = (3.1 \pm 0.1) \cdot 10^{23}$ см$^{-2}$.

---
[3] http://mytorus.com/model-files-mytorus-downloads.html
[4] http://mytorus.com/mytorus-manual-v0p0.pdf

Величины спектральных параметров для наилучших подгонок моделей BNTorus и MYTorus приведены в Табл.1. Изображение спектра для модели MYTorus приведено на Рис. 4 соответственно.

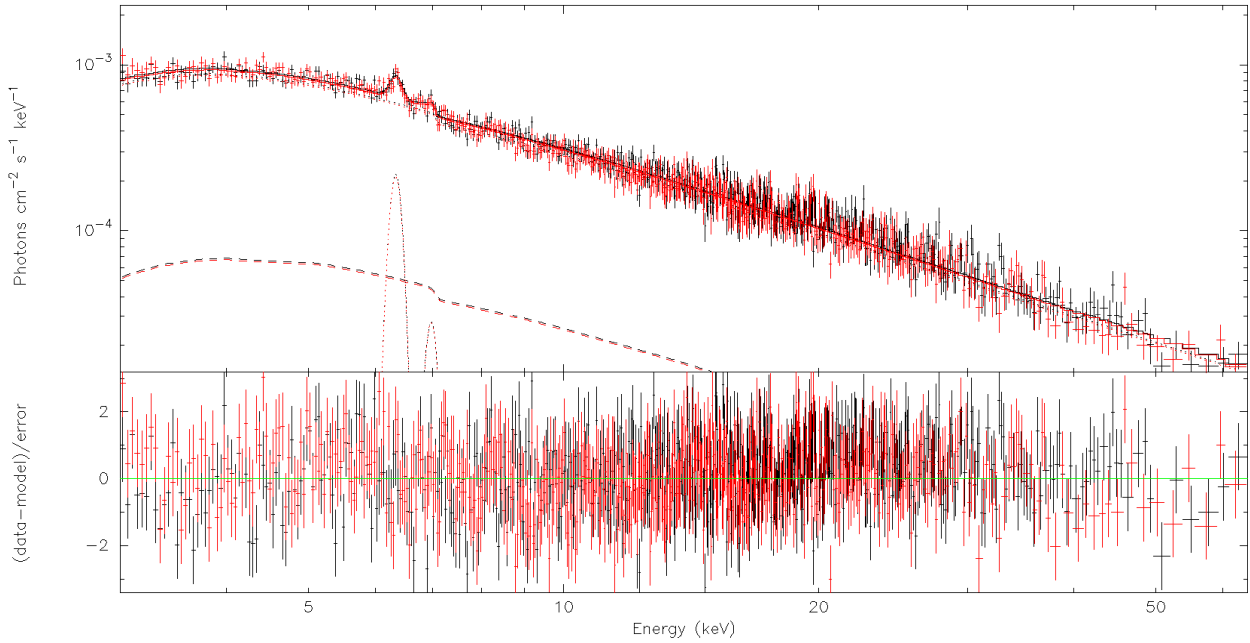

*Рис. 4*. Лучшая подгонка спектра с использованием модели MYTorus. Нижняя панель – остаточные отклонения. Сплошная кривая – суммарная модель, точечный пунктир – отдельные вклады степенного континуума и эмиссионных линий, пунктир – компонента рассеянного континуума.

*Таблица 1*. **Значения спектральных параметров для лучшей подгонки спектра NGC 7172.**

| Базовая модель | $\Gamma$ | $N_H$ (10$^{22}$ см$^{-2}$) | R | $E_{line}$ (кэВ) | $\sigma_{line}$ (эВ) | $\Theta_i$ (град) | $\chi^2$/d.o.f. |
|---|---|---|---|---|---|---|---|
| pexrav | 1.83±0.05 | 8.9±0.5 | 0.50±0.13 | 6.33±0.06 | $93^{+100}_{-90}$ | 60(f) | 1106/1124 |
| pexmon | 1.80±0.02 | 8.6±0.3 | $0.43^{+0.08}_{-0.07}$ | - | $59^{+110}_{-59}$ | 60(f) | 1104/1124 |
| plcabs | 1.79±0.02 | 8.2±0.4 | 0.41±0.05[б] | - | $94^{+111}_{-94}$ | 60(f) | 1109/1124 |
| BNTorus | 1.80±0.01 | 7.7±0.2 | 0.35±0.04[в] | - | 93(f) | $75^{+u}_{-12}$ | 1116/1125 |
| MYTorus | 1.71±0.01 | 31.0±1.0[a] | - | - | 93(f) | $61.1^{+0.9}_{-0.4}$ | 1143/1125 |

Примечания: [a]Приведена величина $N_{H(eq)}$. [б,в]Отражение согласно дополнительной компоненты pexmon.

# 5 Обсуждение и результаты

Мы представляем результаты первого наблюдения обсерваторией *NuSTAR* галактики NGC 7172 типа Сейферт 2. Было проанализировано спектр в диапазоне энергий 3-64 кэВ с помощью как феноменологических моделей, так и численных `BNTorus` и `MYTorus`. Фактически, все предпринятые модели показывают хорошую подгонку, а значения их параметров сопоставимы между собой.

**Континуум.** Значения степенного индекса в зависимости от модели Г~1.71-1.83, сравнимы со значениями, полученными другими авторами с использованием широкодиапазонных данных (например, [7,16,28]). Анализ в диапазоне 0.3-10 кэВ по данным XMM-Newton показывает более плоский спектр с Г~1.55-1.65 (например, [11,13,17]), возможно, по причине влияния компоненты отражения на более высоких энергиях. Присутствие этой компоненты было уставлено в работах с использованием в широком диапазоне энергий данных *BeppoSAX* [7][5], *XMM-Newton+INTEGRAL/ISGRI* [8,36] и *Suzaku* [16]. Небольшие оценки параметра отражения R из последних трёх работ, а именно, R=0.3±0.1[6], R=0.33±0.17 и $R = 0.34^{+0.10}_{-0.09}$ соответственно, совпадают в пределах ошибки с нашей оценкой <R>~0.44[7], что значит, что спектр NGC 7172 не является отражённо-доминирующим.

Полученное значение поглощения $N_H \approx 8 \cdot 10^{22}$ см$^{-2}$ полностью согласуется с результатами всех предыдущих анализов рентгеновских наблюдений, что свидетельствует о неизменном его значении на интервале почти в 30 лет. Отметим, что почти одинаковые значение были получены при подгонке моделями с разной геометрией (т. е. `pexrav` и `MYTorus`). Не смотря на стабильную величину поглощения, наблюдения демонстрируют значительную долговременную вариацию источника более чем на порядок (см. Рис 5, верхняя панель). Анализ представленного наблюдения *NuSTAR* с двумя разными упомянутыми приближениями показывает, что внутренние светимости, исправленные на поглощение, лежат в диапазоне $L_{intr}$(2-10 кэВ)=(1.14-1.23)·10$^{43}$ эрг/с, $L_{intr}$(10-40 кэВ)=(1.56-1.62)·10$^{43}$ эрг/с, что соответствует повышению яркости источника. Для вычисления болометрической светимости воспользуемся фактором ~10, выведенным для диапазона 2-10 кэВ в работе [19], что даёт нам $L_{bol} \approx 1.19 \cdot 10^{44}$ эрг/с. Принимая оценку масс центральной сверхмассивной чёрной дыры (СМЧД) ~4.5·10$^8$ $M_{Sun}$ из наблюдения с высокой разрешающей способностью в ближнем ИК на телескопе VLT (UT4, Yepun) [31], можно вычислить Эддингтоновское отношение $L_{bol}/L_{edd} \approx 1.19 \cdot 10^{44}/5.62 \cdot 10^{46} = 2.1 \cdot 10^{-3}$ (или $\log_{10}(L_{bol}/L_{Edd})$=-2.67). Такой низкий темп аккреции может означать наличие режима аккреции с неэффективно излучающим потоком или RIAF (Radiatively Inefficient Accretion Flow) (см. например, [4,26]). Данный тип аккреции может показывать степенной индекс в диапазоне Γ~1.4-1.9, подразумевает присутствие горячей короны, а также предсказывает возможное присутствие биполярных оттоков. NGC 7172 не является радио громкой галактикой, но в работе [32] при анализе наблюдений VLA на частоте 8.4 ГГц ядра галактики была найдена юго-западная удлинённая структура размером 67 пк, похожая на слабый джет или отток. Поскольку в работах [13,17,16] использовались другие значения фактора перевода в $L_{bol}$ и другое значение массы СМЧД, для сравнения было пересчитано $L_{bol}/L_{edd}$ и получено, что, начиная с наблюдения *XMM-Newton* в 2002 году, Эддингтоновское отношение возрастает с величины ~7·10$^{-3}$ (или $\log_{10}(L_{bol}/L_{Edd})$=-3.16). Интересно отметить, что сценарий с аккрецией в режиме RIAF согласуется с тем, что долгое время NGC7172 относилась к классу галактик без скрытой области широких линий или NHBLR (Non-Hidden Broad-Line Region) [18] и только в работе [31] авторы обнаружили слабые широкие лини Pa$_\alpha$ (1.875 мкм) и Br$_\gamma$ (2.16 мкм), а отсутствие других широких линий (например, H$_\alpha$ и H$_\beta$) они объясняют перекрытием пылевой полосой галактики. С другой стороны, режим RIAF подразумевает геометрически толстый внутренний регион аккреционного диска, который частично может заполнять область BLR, таким образом, уменьшая её объём.

---
[5] Хотя присутствие отражения установлено на уровне более 90%, из-за плохой интеркалибровки между инструментами MECS и PDS, значение параметра отражения вычислено приблизительно (R~1-3).
[6] Отметим, что к значению R, полученному в работе [8], нужно отнестись с осторожностью, поскольку верхний предел для $\Delta E_{cut}^{up}$=56 кэВ близок к области «комптоновского горба» ~20-40 кэВ, что может привести к недооценке/переоценке этих величин.
[7] Без учёта отражения, полученного с моделью `BNTorus`.

**Линия Fe K$_\alpha$.** В соответствии с предыдущими наблюдениями, данные *NuSTAR* также показывают присутствие эмиссионной линии около 6.4 кэВ, энергия которой согласуется с нейтральной линией Fe K$_\alpha$. Измеренная эквивалентная ширина линий равняется $EW = 67^{+13}_{-14}$ эВ, поток в линии $F_{FeK\alpha}$= (1.92±0.43)·10$^{-5}$ фотонов/см$^2$/с или $(1.95^{+0.35}_{-0.56})$·10$^{-13}$ эрг/см$^2$/с.

Для того, чтобы установить место происхождения линии Fe K$_\alpha$, было проведено сравнение полученных в этой работе значений параметров линии $EW_{FeK\alpha}$, $F_{FeK\alpha}$ и континуума $L_{intr}$ с результатами наблюдений за предыдущие ~30 лет. В таблице 2 приведено, а на Рис. 5 отображено эволюцию значения $EW_{FeK\alpha}$ вместе со светимостью $L_{intr}$ в диапазоне 2-10 кэВ, а на Рис. 6 – изменение интенсивности линии вместе с качественными данными наблюдений *XMM-Newton* [17] и *Suzaku* [9]. Данные показывают чёткую антикорреляцию между изменениями $EW_{FeK\alpha}$, $F_{FeK\alpha}$ и $L_{intr}$. Изменение светимости происходит больше, чем на порядок, а $EW_{FeK\alpha}$ приблизительно в ~4 раза. Основной эффект в том, что ширина линии $EW_{FeK\alpha}$ возрастает с уменьшением светимости и наоборот.

Считается, что двумя основными областями генерации линии Fe K$_\alpha$ являются аккреционный диск и газопылевой тор. Если линия генерируется в аккреционном диске, тогда согласно [10], угол его наклона для текущей $EW_{FeK\alpha}$ должен составлять около ~70˚, по результатам работы [9] ~80˚, а согласно наблюдениям *BepoSAX* [7], – быть ориентированным почти плашмя. Более реалистичной причиной вариации $EW_{FeK\alpha}$ может быть изменение состояния короны аккреционного диска – ее температуры или оптической толщи. Но тогда временная задержка изменения $F_{\text{FeK}\alpha}$ линии по отношению к континууму должна составлять от десятков минут до нескольких часов (т.е. быть соизмеримой с размерами аккреционного диска), а результирующая эквивалентная ширина быть неизменной. Но такового не наблюдается.

С другой стороны, применив модели газопылевого тора по данным *NuSTAR*, было найдено, что линия Fe K$_\alpha$ возможно генерируется в поглощающей структуре с фактором раскрытия ~0.5 и величиной поглощения $N_\text{H}$~8·10$^{22}$ см$^{-2}$. Кроме того, применение модели `MYTorus` дало значение $A_S = A_L$, отличное от 1, а именно $A_S = A_L \approx 1.2$, и в общем, следуя [40,42], это может быть интерпретировано как проявление задержки отклика рассеянного континуума на переменность центрального источника, т.е. часовой масштаб отклика больше времени накопления наблюдения. Последнее полностью согласуется с взаимным поведением $EW_{\text{FeK}\alpha}$, $I_{\text{FeK}\alpha}$ и $L_{\text{intr}}$. Используя данные из таблицы 2, при простейшем приближении можно вычислить ориентировочное расстояние между центральным источником и источником линий (т.е. d≈c·Δt, Δt – наблюдаемое время вариации) d~12 лет~3.7 пк. Подчеркнем, что полученная величина расстояния d~3.7 пк совпадает с типичными значениями размеров газопылевого тора.

Прямое сопоставление значения $EW_{\text{FeK}\alpha}$ к измеряемому $N_\text{H}$ или параметру отражения для Сейфертов 2 типа не является полностью корректным из-за присутствия эффекта Балдвина [14] (зависимость $EW_{\text{FeK}\alpha}$ — $L_{\text{intr,x-ray}}$), впервые установленного для данного типа АЯГ в работе [27] с использованием выборки [9]. Согласно результатам в [27], для корректного учёта влияния поглощения на континуум, при допущении образования линии Fe K$_\alpha$, лучше использовать соотношение светимостей $L_{\text{intr}}$(10-50 кэВ) и $L_{\text{FeK}\alpha}$. Наблюдаемая log($L_{\text{FeK}\alpha}$)=41.23. При использовании уравнения 2 в [27] с имеющейся $L_{\text{intr}}$(10-50 кэВ)=1.93·10$^{43}$ эрг/с, ожидаемая log($L_{\text{FeK}\alpha}$)=41.76-39.90, т.е. полностью соответствует измеренной[8]. Для эффекта Балдвина (уравнение 7 в [27] с параметрами из Рис.6(b) в [16]), вычисленный наклон $\omega$ зависимости "log($L_{\text{FeK}\alpha}/L_{\text{intr}}$(10-50 кэВ)) – log($L_{\text{intr}}$(10-50 кэВ))" составляет ~0.05, что меньше полученного в [27], но сопоставимо с таким в [16]. Объяснение расхождения может заключаться в том, что в [16] используется более однородная выборка галактик только с диапазоном 22≤log($N_\text{H}$)<24 и без радио-громких источников. Таким образом, вариант, в котором газопылевой тор есть источником линии Fe K$_\alpha$, лучше описывает спектральные и временные характеристики рентгеновского излучения активного ядра в NGC 7172.

К сожалению, изучить поведение параметра отражения R со временем не представляется возможным по причине наличия только двух оценок данного параметра с небольшими ошибками.

Исходя от полученных результатов анализа для линии Fe K$_\alpha$, отметим, что заключение о её образовании в поглощающей среде согласуется с таким в работе [7]. При этом, оно противоречит заключению о формировании в аккреционном диске [1], в основном, из-за наличия лучшего качества данных и обнаружения большой по времени задержки в переменности $EW_{FeK\alpha}$ и $L_{intr}$.

---
[8] С целью упрощения и возможности сравнения значений параметров с работами других авторов, использованные здесь величины log($L_{\text{FeK}\alpha}$) и $L_{\text{intr}}$(10-50 кэВ) были вычислены в результате базового моделирования на основе модели `pexrav`.

*Таблица 2.* **Величины собственной светимости в диапазоне 2-10 кэВ и эквивалентной ширины линии Fe $K_\alpha$, использованные для графика на Рис.5.**

| Дата наблюдения | Инструмент | $L_{intr}$ (эрг/сек) | $EW_{FeK\alpha}$ (эВ) | Статьи |
|---|---|---|---|---|
| 28.10.1985 | *EXOSAT* | $2.10 \cdot 10^{43}$ | - | [33] |
| 26.10.1989 | *Ginga* | $1.10 \cdot 10^{43}$ | 71±39 | [24] |
| 12.05.1995 | *ASCA* | $1.65 \cdot 10^{43}$ | 68±35 | [34] |
| 15.10.1996 | *BeppoSAX* | $3.50 \cdot 10^{42}$ | 120±65 | [1,7] |
| 06.11.1997 | *BeppoSAX* | $1.70 \cdot 10^{42}$ | 210±105 | [1,7] |
| 18.11.2002 | *XMM-Newton* | $3.98 \cdot 10^{42}$ | 200±50 | [17] |
| 11.11.2004 | *XMM-Newton* | $5.01 \cdot 10^{42}$ | 120±10 | [17] |
| 24.04.2007 | *XMM-Newton* | $7.94 \cdot 10^{42}$ | 100±20 | [17] |
| 25.05.2008 | *Suzaku* | $1.00 \cdot 10^{43}$ | 52±4 | [16] |
| 07.10.2014 | *NuSTAR* | $1.14 \cdot 10^{43}$ | 67±14 | эта работа |

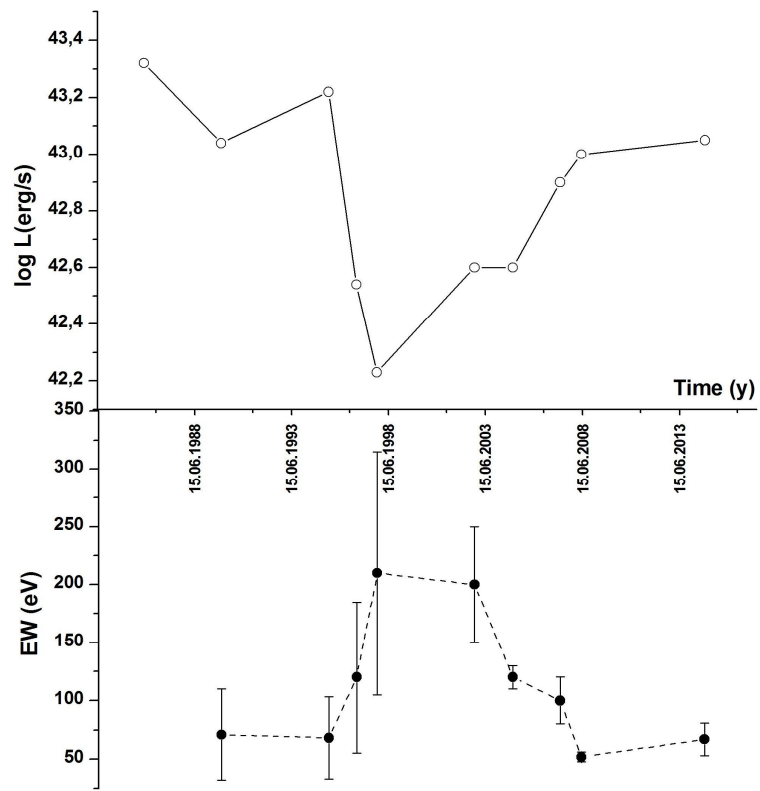

*Рис. 5.* Изменение собственной светимости активного ядра галактики NGC 7172 в диапазоне 2-10 кэВ и EW Fe $K_\alpha$.

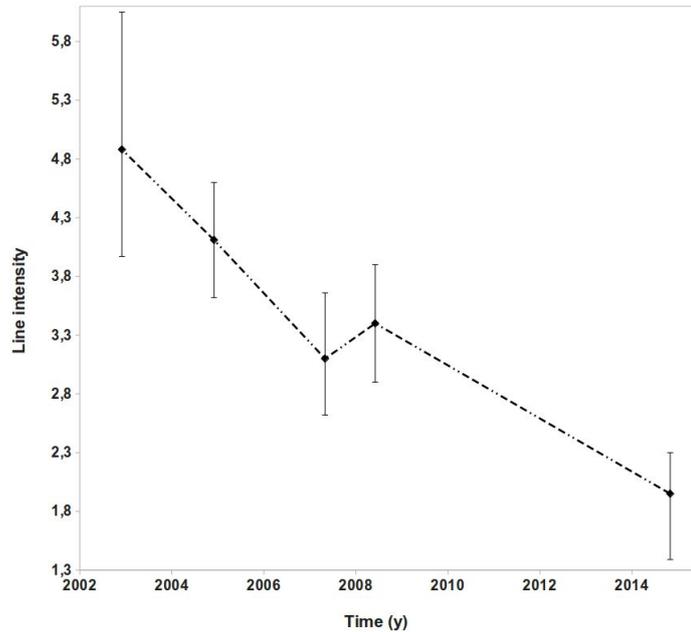

*Рис. 6*. Изменение интенсивности лини Fe $K_\alpha$ по данным *XMM-Newton*, *Suzaku* и *NuSTAR*. Поток в линии в единицах $10^{-13}$ эрг/см$^2$/с.

В перспективе, проведение широкодиапазонных рентгеновских наблюдений NGC 7172 с использованием миссий *NuSTAR*, *XMM-Newton* или будущей миссии *Athena* даст возможность детальнее изучить спектр отражения, попытаться разделить в нем вклады от аккреционного диска и газопылевого тора, а также детальнее изучить переменность объекта на долгих промежутках времени.

**Благодарности**